\documentclass[12pt]{article}
\usepackage[margin=1in]{geometry}
\usepackage{amsmath,amssymb,amsthm}
\usepackage{mathtools}
\usepackage{booktabs}
\usepackage{enumitem}
\usepackage{natbib}
\usepackage{hyperref}
\usepackage{fancyhdr}
\usepackage{xcolor}
\usepackage{graphicx}
\usepackage{float}

\newtheorem{proposition}{Proposition}
\newtheorem{remark}{Remark}

\title{\textbf{The Adversarial Discount} \\ \large AI, Signal Correlation, and the Cybersecurity Arms Race}
\author{James Bono}

\begin{document}
\maketitle

\begin{abstract}
We study a contest-theoretic model of adversarial investment in which an attacker and a defender allocate resources to AI-augmented capabilities across multiple attack surfaces. The attacker's investment operates through two channels: it amplifies offensive potency unconditionally and erodes defensive effectiveness conditionally, generating an \emph{adversarial discount} that deepens endogenously with the defender's own investment. We derive a closed-form \emph{arms race ratio} decomposing the relative marginal effectiveness of offensive and defensive investment into six structural primitives, and establish equilibrium uniqueness and global convergence under a continuous best-response dynamic. The central result concerns signal cross-correlation, the degree to which threat intelligence on one surface informs detection on another. With full cross-correlation, the arms race ratio is independent of the number of attack surfaces: the attacker's structural advantage from surface proliferation is completely neutralized. Under the benchmark full-dilution case, without cross-correlation, per-surface defense effectiveness vanishes as the attack surface grows. Extending the analysis to heterogeneous defenders facing an attacker who targets by expected value, we argue that the model points to a dual inefficiency: overinvestment in private defense (a zero-sum redirective externality) and underinvestment in shared signal correlation (a public good). These formal results, together with public-good reasoning outside the base model, characterize when collective information aggregation can dominate private capability investment as the decisive margin in adversarial contests.

\medskip\noindent\textbf{Keywords:} cybersecurity economics, AI arms race, adversarial discount, signal correlation, platform security, game theory
\end{abstract}

\section{Introduction}
\label{sec:Intro}

The cost of AI-augmented attack tooling has fallen dramatically. Open-source models, AI-as-a-service APIs, and commoditized attack frameworks have placed offensive AI capabilities within reach of the full spectrum of threat actors. The question facing defenders is no longer whether to invest in AI-enabled security, but how to invest, and along which margins investment is most productive.

This paper develops a game-theoretic model of the attacker-defender arms race organized around two ideas.

The first is the \emph{adversarial discount}: a structural property in which defender AI effectiveness per unit of investment declines as the attacker's AI adapts. The mechanism is not reducible to any particular technology failure, such as pattern-trained classifiers degrading against novel inputs. It is more fundamental. The attacker \emph{optimally directs} AI investment against the defender's deployed capabilities. If the defender excels at detection, the attacker invests in evasion; if the defender excels at triage, the attacker floods the zone. The discount is a consequence of the adversarial game itself, and we model it through a continuous function that declines in attacker AI investment. The result is a dual channel: attacker AI simultaneously amplifies offensive potency and erodes defensive effectiveness.

The adversarial discount is not new to AI. The \emph{assume breach} posture in security practice reflects the same structural property: given sufficient time and resources, a determined attacker finds a way in, because adversarial innovation responds to every defense deployed. What AI changes is the \emph{economy} of offense. Stages of attack that were previously limited by economies of skill, including vulnerability discovery, exploit development, lateral movement, and credential harvesting, now benefit from economies of scale in compute. The discount deepens endogenously with the defender's own investment not because AI is uniquely dangerous, but because AI sharply lowers the cost of the adversarial adaptation that generates it.

A natural question is whether defenders can neutralize the discount by using AI to attack themselves: running automated red-team tools to find and patch their own vulnerabilities. Defensive self-attack can reduce the magnitude of the discount, but not eliminate it. Attackers adapt in real time to deployed defenses, but defenders must convert discovered vulnerabilities into safe system changes subject to operational risk, uptime requirements, compatibility, and coverage across many surfaces. Those differences in objectives mean the discount is structural, not technological.

The second is \emph{signal cross-correlation}: the degree to which threat intelligence from one attack surface informs detection on another. We show that signal cross-correlation is the decisive variable in the arms race. With full cross-correlation, the attacker's structural advantage from surface proliferation is completely neutralized, and defense remains effective regardless of how many surfaces the defender must protect. Without cross-correlation and under signal dilution ($\rho > 0$), the defender's investment is spread across surfaces and can become irrelevant at enterprise scale, with breach probability converging toward the level implied by no effective defense at all. Population-level signal cross-correlation, however, is a public good. Individual defenders can correlate across their own surfaces, but detecting campaign-level patterns requires signals from other defenders' environments that no single organization possesses.

The adversarial discount also offers a historical interpretation of how the AI arms race began. When AI adoption costs were high, attackers preferred conventional methods, but defensive AI investment created a structural opportunity by providing more deployed capability to erode. Above a threshold $\hat{d}$, the erosion channel made AI-augmented attacks profitable even with fixed adoption costs $F > 0$. This \emph{provocation mechanism} is now historically descriptive rather than strategically prescriptive. With $F \approx 0$ today, attackers use AI regardless of $d$, and the relevant question has shifted from whether to invest to how the arms race evolves.

Finally, we step outside the formal model to develop economic arguments that extend its logic. When the attacker faces a population of defenders and targets the highest expected-value combination of vulnerability and asset value, we argue that the resulting equilibrium exhibits a dual inefficiency: simultaneous overinvestment in private defense, which is a zero-sum positioning game, and underinvestment in shared signal cross-correlation, which is a positive-sum public good. This conjecture follows from the model's structure under standard public-good reasoning, though it has not been formally derived in a $K$-defender equilibrium. It connects to the longstanding policy question in cybersecurity of how to sustain collective action in threat-intelligence sharing.

The paper proceeds as follows. Section \ref{sec:Lit} situates our contribution in the literature. Section \ref{sec:model} presents the model and derives the single-surface arms race ratio, establishes equilibrium uniqueness and global convergence. Section \ref{sec:signal} develops the central contribution: signal cross-correlation and multi-surface defense, including the multi-surface arms race ratio and scaling results. Section \ref{sec:Extensions} extends the model's insights (but not its formalism) to heterogeneous defenders and platform economics.

\section{Related Work and Contribution}
\label{sec:Lit}

This paper relates to the literatures on cybersecurity investment, network security games, contest theory, and AI in cybersecurity. Our contribution is to introduce signal correlation as a central determinant of equilibrium outcomes in attacker–defender settings.

\paragraph{Cybersecurity investment economics.} The canonical model is \citet{gordon2002economics}, which characterizes optimal security investment for a single defender facing an exogenous breach function and implies the $1/e$ upper bound on investment. Our model nests Gordon-Loeb in the absence of attacker response. Introducing a strategic attacker alters the effective breach function: attacker investment erodes defender effectiveness, reducing concavity and increasing optimal investment. This is consistent with \citet{ebel2024economics}, who study a two-sided Stackelberg game. Our contribution is to introduce the adversarial discount as a structural erosion channel and to characterize its implications for equilibrium investment, including the arms race ratio and signal correlation effects.

\citet{gordon2015externalities} incorporate externalities and show that social investment exceeds private investment under contagion. In contrast, we show that under strategic targeting, private incentives can be distorted in both directions: overinvestment in private defense (due to attack redirection) and underinvestment in shared correlation. This connects to the literature on information sharing in cybersecurity \citep{pala2019information}, which emphasizes the public-good nature of shared defensive information.

\paragraph{Network security and strategic interaction.} \citet{acemoglu2016network} develop a general framework for security investment in networks, showing that strategic substitutability and targeting can reverse standard underinvestment results. Our population-level analysis builds on their own-effect/externality decomposition to show that different dimensions of investment (private defense vs.\ signal correlation) can generate externalities of opposite sign. However, their model of network contagion implicitly assumes that network neighbors of a compromised node face the same effective security against propagation as against a direct attack. Supply-chain compromises of the SolarWinds type break this assumption. When an upstream supplier is compromised, downstream propagation is automatic and conditional on that compromise. Importantly, in the supply chain attack setting, the signal correlation mechanism analyzed here cannot operate against the compromise itself, only against the attacker's subsequent exploitation to inflict damages. This limits the reach of signal correlation in supply-chain threat models: it governs the defender's detection advantage during exploitation, not during the initial compromise propagation.

\citet{goyal2014attack} study attack, defense, and contagion in network settings, while \citet{roy2010survey} and \citet{hunt2024review} provide surveys of game-theoretic approaches to network security and attacker--defender models more broadly. None of these papers models AI-driven erosion of defensive effectiveness or cross-surface signal correlation.

\paragraph{Contest theory.} Our breach function is a Tullock contest success function, grounded in the axiomatic characterization of \citet{skaperdas1996contest}. We depart from the standard formulation by allowing attacker investment to endogenously degrade the defender’s effectiveness parameter. This preserves the contest structure while introducing an erosion channel that generates the arms race ratio and convergence dynamics. \citet{iliaev2022tullock} apply contest models to multi-asset cybersecurity settings but do not study AI-driven dynamics or signal correlation.

\paragraph{AI and cybersecurity.} \citet{Garg2024artificial} and \citet{heitzenrater2025winning} argue that AI may reduce attacker advantage under appropriate conditions. Our model provides a formalization of this hypothesis: when adversarial erosion is limited and signal correlation is high, defender effectiveness improves. \citet{hausken2024review} surveys attacker--defender models in cybersecurity, including dynamic and incomplete-information settings, but does not consider AI-specific erosion or cross-surface signal aggregation.

\paragraph{Contribution.} Our main contributions are as follows. We introduce the adversarial discount, a mechanism through which attacker effort endogenously erodes defender effectiveness, and derive the arms race ratio that characterizes the relative marginal effect of offensive and defensive effort in terms of six structural primitives. We then model cross-surface signal correlation and establish that it can fully offset the scaling disadvantage of attack surface expansion. Finally, we argue that these mechanisms generate a dual inefficiency in multi-defender settings: excessive private defense effort alongside insufficient provision of shared signal infrastructure. The unifying theme is that the decisive margin in AI-era cybersecurity is not the scale of capability investment but the structure of information aggregation.

\section{Model and Baseline}\label{sec:model}

\subsection{Setup}

Two players: a Defender~(D) and an Attacker~(A). D chooses AI investment $d \geq 0$ at cost $c_d \cdot d$, and A chooses AI investment $a \geq 0$ at cost $c_a \cdot a + F \cdot \mathbf{1}(a > 0)$, where $F \geq 0$ is a fixed cost of AI adoption.

The breach probability on a single surface takes a Tullock contest form:
\begin{equation}\label{eq:breach}
q(a, d, s) = \frac{q_0 \cdot h(a)}{q_0 \cdot h(a) + (1 - q_0)\bigl(1 + \delta(a) \cdot d \cdot s\bigr)}
\end{equation}
guaranteeing $q \in [0,1]$ for all parameter values. At the status quo $(a=0, d=0)$, $q = q_0$.

The parameters: $q_0 \in (0,1)$ is the baseline breach probability, representing attacker success under conventional non-AI effort at the status quo. $h\colon \mathbb{R}_+ \to [1,\infty)$ is the \emph{attack amplification function} ($h(0)=1$, $h'>0$, $h''<0$), capturing how attacker AI increases breach capability unconditionally. $\delta\colon \mathbb{R}_+ \to (0, \delta_0]$ is the \emph{defender effectiveness function} ($\delta(0) = \delta_0$, $\delta' < 0$, $\delta'' > 0$), capturing how attacker AI erodes defender effectiveness conditional on the defender having invested. The convexity $\delta'' > 0$ means that the absolute erosion $\lvert\delta'\rvert$ is diminishing: each additional unit of attacker AI erodes less defense than the previous unit. $s \geq 0$ scales the effectiveness of defender AI investment and acquires a structural signal interpretation in the multi-surface extension (Section \ref{sec:signal}).

The two functions are conceptually distinct. Amplification $h(a)$ operates unconditionally: at $d=0$, breach probability increases in $h$ regardless of defender behavior. Erosion through $\delta(a)$ is conditional on $d > 0$, entering through the term $\delta(a) \cdot d \cdot s$ and vanishing when the defender has not invested. The distinction matters because collapsing both into a single function would make the attacker's marginal return independent of the defender's investment level, eliminating the strategic interaction between the two sides' AI investments.

Define the \emph{adversarial discount} $1 - D(a)$, where $D(a) = \delta(a)/\delta_0 \in (0,1]$ is the fraction of defender effectiveness retained. Define the \emph{adversarial leverage} $h(a)/\delta(a)$, which is strictly increasing in $a$.

Payoffs: $U_D = -V \cdot q - c_d \cdot d$ and $U_A = B \cdot q - c_a \cdot a - F \cdot \mathbf{1}(a > 0)$.

\noindent\textit{Units.} Since $\delta \cdot d \cdot s$ is added to~1 inside the contest form, the product $\delta ds$ must be dimensionless. Accordingly, $d$ and $a$ are dimensionless intensities of AI deployment, not dollar amounts; the mapping from budget to deployment intensity involves a production function that absorbs the units. This is standard in contest models; see \citet{skaperdas1996contest} for the axiomatic characterization of effort in contest success functions.

Writing $\Phi = q_0 h(a) + (1-q_0)(1 + \delta(a)ds)$ for the contest denominator, the attacker's first-order condition is:
\begin{equation}\label{eq:foc}
B \cdot \frac{q_0(1-q_0)\bigl[h'(a)(1+\delta(a)ds) + h(a)\lvert\delta'(a)\rvert ds\bigr]}{\Phi^2} = c_a
\end{equation}
The bracketed term decomposes into the amplification channel ($h'$, present even at $d=0$) and the erosion channel ($h \lvert\delta'\rvert ds$, proportional to $d$). The erosion channel creates a strategic complementarity between the two sides' investments. As the defender invests more, the erosion term grows, raising the attacker's marginal return to AI investment. The defender's own investment feeds the attacker's incentive to invest, not because attacker AI becomes intrinsically more potent, but because there is more deployed defense to erode.

\subsection{The Single-Surface Arms Race Ratio}\label{sec:arms_race}

We define the \emph{arms race ratio} $R$ as the ratio of the attacker's marginal effectiveness to the defender's. When $R < 1$, each unit of defensive AI reduces insecurity more than each unit of offensive AI increases it; when $R > 1$, the attacker's investment is more effective. Taken over all values of $(d,a)$, $R$ characterizes the relative effectiveness of the two sides across the investment space. $R$ is a local diagnostic of relative effectiveness; equilibrium outcomes depend on best responses, costs, and how the system evolves along the adjustment path.

On a single surface, $R = (\partial q/\partial a)/(-\partial q/\partial d)$. A useful property of the contest form is that the denominator $\Phi^2$ appears in both the numerator and denominator of this ratio and cancels, so $R$ has the same expression regardless of whether $q$ is large or small. At the status quo ($d = 0$, $a = 0$), only the amplification channel is active because the erosion channel is proportional to $d$ and thus vanishes. This gives:
\[
R_0^{\text{single}} = \frac{\alpha}{\delta_0 \cdot s}
\]
where $\alpha \equiv h'(0)$ is the marginal AI effectiveness for attack. When the arms race is underway ($d > 0$, $a > 0$), the erosion channel activates and the general expression is:
\begin{equation}\label{eq:Rgeneral}
R(d,a) = \underbrace{\frac{h'(a)(1 + \delta(a)\,d\,s)}{h(a)\,\delta(a)\,s}}_{\text{amplification}} + \underbrace{\frac{\lvert\delta'(a)\rvert\,d}{\delta(a)}}_{\text{erosion premium}}
\end{equation}
The first term captures the attacker's marginal return from generating more potent attacks. The second term, the \emph{erosion premium}, is the attacker's additional return from degrading the defender's deployed capabilities. The erosion premium is zero at $d=0$ and proportional to $d$, so the defender's own investment feeds the attacker's marginal return. Since both terms are increasing in $d$ at any fixed $a$, higher defender investment raises $R$, and the defender's initial advantage ($R_0 < 1$) can erode as its own investment grows.

\subsection{Equilibrium Uniqueness and Dynamics}
\label{sec:dynamics}

The results above characterize the arms race ratio at any given $(d, a)$. We now ask whether the arms race settles at all, and if so, whether the outcome is unique. To address this, we adopt a Nash equilibrium framework.

\paragraph{Analytical structure.} The contest form provides more analytical tractability than might be expected from the implicit best-response functions. At any interior defender optimum, the FOC $V \cdot q_0(1-q_0)h\delta s/\Phi^2 = c_d$ pins down the contest denominator:
\begin{equation}\label{eq:phi_closedform}
\Phi^2 = \frac{V\, q_0(1-q_0)\, h(a)\,\delta(a)\, s}{c_d}
\end{equation}
which is a function of $a$ alone (the defender's investment $d$ drops out). This yields the defender's interior best response in closed form:
\[
d^*(a) = \frac{\Phi(a) - q_0 h(a) - (1-q_0)}{(1-q_0)\,\delta(a)\,s}
\]
and the corresponding breach probability along the interior branch is $q^*(a) = q_0 h(a)/\Phi(a) = \sqrt{q_0 c_d\, h(a)\,/\,(V(1-q_0)\delta(a)s)}$, which is increasing in the adversarial leverage $h/\delta$.

Substituting the interior defender best response $d^*(a)$ into the attacker's payoff reduces the two-dimensional fixed-point problem to a single-variable optimization: the attacker maximizes $B \cdot q^*(a) - c_a \cdot a$, which is $B\sqrt{q_0 c_d \cdot h(a)/\delta(a)\,/\,(V(1-q_0)s)} - c_a a$. If $\sqrt{h(a)/\delta(a)}$ is strictly concave, this objective is strictly concave in $a$, giving a unique interior attacker optimum $a^*$ and therefore at most one interior equilibrium $(d^*(a^*), a^*)$.

\begin{proposition}[Interior Equilibrium Uniqueness]\label{prop:uniqueness}
In the single-surface contest-form model, if an interior equilibrium exists and $\sqrt{h(a)/\delta(a)}$ is strictly concave, then that interior equilibrium is unique. Equivalently, there is at most one interior equilibrium. A sufficient condition is $h'' < 0$ (concave amplification) together with $\delta \cdot \delta'' \geq 2(\delta')^2$ (non-accelerating erosion rate). For the hyperbolic form $\delta(a) = \delta_0/(1+\beta a)$, the second condition holds with equality, and $h'' < 0$ alone guarantees interior uniqueness.
\end{proposition}

The condition $\delta \cdot \delta'' \geq 2(\delta')^2$ restricts how fast the adversarial discount deepens, requiring that the erosion function cannot decline faster than $1/(1+\beta a)$. For power-law forms $\delta = \delta_0/(1+\beta a)^k$, the condition holds if and only if $k \leq 1$, ruling out erosion functions that are more concave than the hyperbolic form. Models with infinitely fragile defender technology (e.g., exponential erosion $\delta = \delta_0 e^{-\beta a}$) can generate multiple equilibria and are excluded.

\paragraph{Global convergence.} Let $d^{BR}(a) = \max\{0, d^*(a)\}$ denote the defender's constrained best response, and let $a^{BR}(d)$ denote the attacker's constrained best response. In the continuous-time limit ($\eta \to 0$, rescaling time), the dynamics become:
\[
\dot{d} = d^{BR}(a) - d, \qquad \dot{a} = a^{BR}(d) - a
\]
\begin{proposition}[Global Convergence]\label{prop:stability}
If the constrained best responses are bounded and the continuous-time best-response dynamics admit a unique equilibrium, then they converge globally to that equilibrium from any initial condition.
\end{proposition}

Under the conditions of Proposition~\ref{prop:stability}, the single-surface arms race therefore settles from any starting point. Regardless of whether both sides begin with heavy AI investment, light investment, or asymmetric positions, the system converges to the same steady state whenever the equilibrium is unique. The substantive question is where that steady state lies, which depends on the depth of the adversarial discount relative to the defender's signal capacity. Section~\ref{sec:signal} introduces the multi-surface setting where cross-correlation $\gamma$ enters as the decisive determinant.

\begin{figure}[H]
\centering
\includegraphics[width=\textwidth]{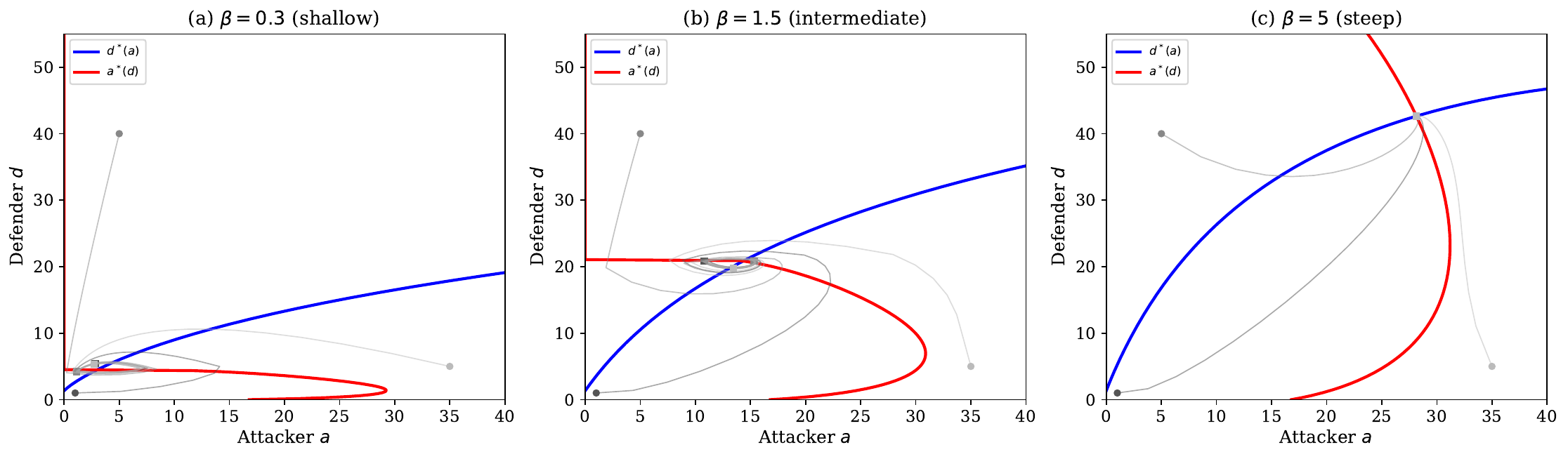}
\caption{Best-response functions and dynamic adjustment paths for three discount steepness values (single surface). Blue curves show the defender's best response $d^*(a)$, which is always increasing (closed-form via equation~\ref{eq:phi_closedform}). Red curves show the attacker's best response $a^*(d)$, which is non-monotone, rising as the erosion channel activates and then falling as the contest structure overwhelms. Gray paths show convergence from different starting points.}
\label{fig:bestresponse}
\end{figure}

For discrete-time dynamics with positive $\eta$, the convergence result extends provided $\eta$ is small enough to prevent overshooting. The character of convergence varies with the adversarial discount. We illustrate this in Figure~\ref{fig:bestresponse} using the hyperbolic form $\delta(a) = \delta_0/(1+\beta a)$.

When the discount is shallow ($\beta$ small), defense is highly effective and the attacker's best response drops to zero at moderate $d$, as AI attacks become unprofitable against a capable defender. The system converges quickly to a defender-favorable equilibrium. At intermediate values of $\beta$, both sides invest at interior solutions. The attacker's best response is non-monotone in $d$, rising initially as the erosion channel makes AI more valuable (there is more defense to erode) and then falling once the contest denominator overwhelms the erosion benefit. Convergence is slower and may involve overshooting, but the system still reaches equilibrium. When the discount is steep ($\beta$ large), the erosion channel is potent enough that the attacker remains committed at high investment levels. Both sides invest heavily and convergence is monotone, but the equilibrium features high $R$. The arms race settles into a costly steady state.

The erosion premium creates a \emph{transitional} cost to delayed investment that is distinct from the long-run equilibrium. Because $R(d,a)$ is increasing in~$d$ (equation~\ref{eq:Rgeneral}), the arms race ratio becomes less favorable to the defender at higher investment levels, a property already internalized by the best-response functions. Propositions~\ref{prop:uniqueness} and~\ref{prop:stability} imply convergence to the same steady state from any initial condition, so early versus late investment does not alter long-run outcomes. The difference is in the adjustment path: a defender that delays investment faces a higher breach probability during the transition and therefore accumulates greater expected losses before reaching steady state. The implication is one of transitional welfare, not equilibrium selection.

\section{Signal Cross-Correlation and Multi-Surface Defense}\label{sec:signal}

\subsection{Signal as a Strategic Variable}

In the single-surface model, $s$ is merely a scalar multiplier on $d$ and is not separately identified. With multiple surfaces, $s$ becomes a budget allocated across surfaces, and the question of cross-surface information flow gives $s$ a structural role that $d$ alone does not carry.

We model \emph{signal dilution with parameter $\rho \in [0,1]$}: $s_i = s/N^\rho$. This nests two limiting cases. When $\rho = 1$ (full dilution), the binding constraint is a fixed signal-processing budget that divides equally across surfaces. When $\rho = 0$ (no dilution), signal-processing capacity scales perfectly with $N$. The intermediate case $\rho \in (0,1)$ captures partial returns to scale in signal processing, in which each additional surface reduces per-surface capacity but less than proportionally. The appropriate value of $\rho$ depends on the nature of the processing bottleneck. When the constraint is analyst attention or SOC queue capacity, which are fixed organizational resources that do not grow with $N$, $\rho$ is close to 1. When the signal-processing infrastructure exhibits near-constant returns to scale (e.g., automated pipelines whose marginal cost per additional surface is small), $\rho$ is closer to 0. We develop the general case and highlight the $\rho = 1$ regime as the conservative benchmark.

Raw telemetry does not dilute (adding a new monitored surface does not reduce logs from existing surfaces), but the capacity to process, correlate, and act on signals is the binding constraint. Cross-correlated signals ($\gamma > 0$) yield effective per-surface signal $s_i^e = s_i + \gamma \sum_{j \neq i} s_j$. With symmetric surfaces, this simplifies to $s_i^e = (s/N^\rho)[1 + \gamma(N-1)]$. The overall breach probability is $P = 1 - \exp(-\lambda)$, where $\lambda = -\sum_i \log(1-q_i)$ is the \emph{log-breach rate}, which is additive and well-defined since $q_i \in [0,1)$ by the contest structure.

Throughout this section, $\gamma$ is treated as an exogenous reduced-form parameter that captures the defender's architectural choices made prior to the AI investment decision. In practice, some determinants of $\gamma$ (data model architecture, product integration, cross-surface detection graphs) are influenced by $d$ investment, but other determinants (pre-existing infrastructure, organizational structure, data governance) are independent. We do not model the cost function of producing $\gamma$; we treat it as a separate strategic variable to isolate the impact of signal cross-correlation on equilibrium outcomes. This reduced-form treatment is sufficient to characterize the central result, namely that $\gamma = 1$ neutralizes the surface-count effect, and to identify signal cross-correlation as a decisive lever. A richer model would endogenize $\gamma$ through a cost function $c(\gamma)$ and allow the defender to jointly choose $(d, \gamma)$, but such an extension is beyond the scope of this paper.

The distinction between within-organization $\gamma$ (correlating signals across one defender's $N$ surfaces) and cross-organization $\gamma$ (correlating signals across many defenders' environments) becomes important in Section~\ref{sec:Extensions}. Individual defenders can raise within-org $\gamma$ through $d$ investment, but cross-org $\gamma$ requires signals from other defenders' environments and cannot be produced individually.

We assume symmetric attack surfaces for tractability. With heterogeneity, two opposing forces arise: (i) concentration of defense on high-value surfaces (reducing effective
$N$), and (ii) attacker targeting of weak links (preserving weakest-link structure). Our results should be interpreted as characterizing the latter force.

\subsection{The Multi-Surface Arms Race Ratio}

With $N$ symmetric surfaces, signal dilution parameter $\rho$, and effective per-surface signal $s_i^e = (s/N^\rho)[1 + \gamma(N-1)]$, the single-surface arms race ratio generalizes. The $\Phi^2$ cancellation still holds per-surface, and the $N/(1-q)$ factors in $\partial\lambda/\partial a$ and $\partial\lambda/\partial d$ cancel in the ratio, so $R$ retains the same form as equation~(\ref{eq:Rgeneral}) with $s$ replaced by $s_i^e$. At the status quo:
\begin{equation}\label{eq:R0}
\boxed{R_0 = \frac{\alpha \cdot N^\rho}{\delta_0 \cdot s \cdot (1 + \gamma(N-1))}}
\end{equation}
The attacker's advantage scales with $N^\rho$ (more targets to probe, modulated by dilution); the defender's advantage scales with $s$ amplified by signal cross-correlation. The general $R(d,a)$ has the same form as equation~(\ref{eq:Rgeneral}) with $s$ replaced by $s_i^e$. The erosion premium is unchanged: it depends on the adversarial discount, not on the number of surfaces or the dilution parameter.

Each parameter in $R_0$ maps to a strategic lever with the following comparative statics:

\begin{center}
\begin{tabular}{llcl}
\toprule
\textbf{Parameter} & \textbf{Meaning} & $\partial R_0 / \partial (\cdot)$ & \textbf{Policy lever} \\
\midrule
$\alpha$ & Attacker AI potency & $+$ & Frontier AI capabilities (exogenous) \\
$N$ & Attack surface count & $+$ at $\gamma < 1$ & IT complexity; surface reduction \\
$\delta_0$ & Defender AI effectiveness & $-$ & Security AI R\&D \\
$s$ & Signal breadth & $-$ & Platform telemetry coverage \\
$\gamma$ & Signal cross-correlation & $-$ & Data integration, detection graphs \\
$\rho$ & Signal-processing returns & $+$ & Technology-dependent (structural) \\
\bottomrule
\end{tabular}
\end{center}

\noindent The defender controls four of the six quantities. The attacker controls $\alpha$ (driven by frontier AI progress), while $\rho$ is a structural property of the signal-processing technology rather than a direct choice variable. The defender's strategic imperative is to invest along the dimensions that keep $R < 1$.

\subsection{Surface Scaling}

\begin{proposition}[Surface Scaling]\label{prop:scaling}
Under signal dilution with parameter $\rho \in [0,1]$:
\begin{enumerate}[label=(\roman*)]
\item When $\gamma = 1$, every surface's effective signal is $s_i^e = sN^{1-\rho}$, and the status-quo arms race ratio is
\[
R_0(\gamma=1) = \frac{\alpha \cdot N^{\rho-1}}{\delta_0 \cdot s}.
\]
At $\rho = 1$, $R_0 = \alpha/(\delta_0 s)$, independent of $N$. For $\rho < 1$, $R_0$ is proportional to $N^{\rho-1}$ and therefore decreasing in $N$.

\item When $\gamma = 0$, each surface sees $s_i^e = s/N^\rho$. For $\rho > 0$, as $N \to \infty$ the defense term vanishes and per-surface breach probability converges to
\[
q_\infty = \frac{q_0\, h(a)}{q_0\, h(a) + (1-q_0)},
\]
which contains no defense terms. The log-breach rate is asymptotically $\lambda \to -N\log(1-q_\infty)$, linear in $N$.

\item The gap $R_0(\gamma=0) - R_0(\gamma=1)$ is increasing in $\rho$. When a critical surface count $N^*$ solving $R_0 = 1$ exists, it is defined implicitly by
\[
N^* = \left(\frac{\delta_0 s [1 + \gamma(N^*-1)]}{\alpha}\right)^{1/\rho}
\]
and, whenever $\alpha \rho (N^*)^{\rho-1} > \delta_0 s\,\gamma$, it is increasing in $\gamma$.
\end{enumerate}
\end{proposition}

Part~(i) says that full signal cross-correlation reduces the surface-count dependence of $R_0$ from $N^\rho$ to $N^{\rho-1}$. At $\rho = 1$, the reduction is complete: $N$ drops out entirely, so the marginal unit of $d$ has the same effect on breach probability relative to the marginal unit of $a$ regardless of how many surfaces the defender must protect. At $\rho < 1$, the reduction is partial but $R_0$ still decreases rather than increases in $N$. At $\rho = 0$ (no dilution), $R_0 = \alpha/(N \delta_0 s)$, decreasing linearly in surface count.

Part~(ii) establishes the polar case. Without cross-correlation and with $\rho > 0$, per-surface defense effectiveness vanishes as $N$ grows, and the log-breach rate converges to a coefficient that reflects no effective defense. This is a limiting result of the dilution assumption, not a prediction that defense is literally useless. At $\rho = 0$, defense does not vanish even without cross-correlation.

Part~(iii) formalizes the interaction between $\gamma$ and $\rho$. When $\rho = 1$, signal cross-correlation is the difference between $R_0$ independent of $N$ and $R_0$ linear in $N$. When $\rho = 0$, cross-correlation still reduces $R_0$ by a factor of $N$, but the baseline $R_0(\gamma=0)$ is already decreasing in $N$, so the marginal value of correlation is lower. The point is not that $\gamma$ inhabits either extreme in practice, but that small increases in $\gamma$ can sharply attenuate the effect of $N$.

\begin{figure}[H]
\centering
\includegraphics[width=\textwidth]{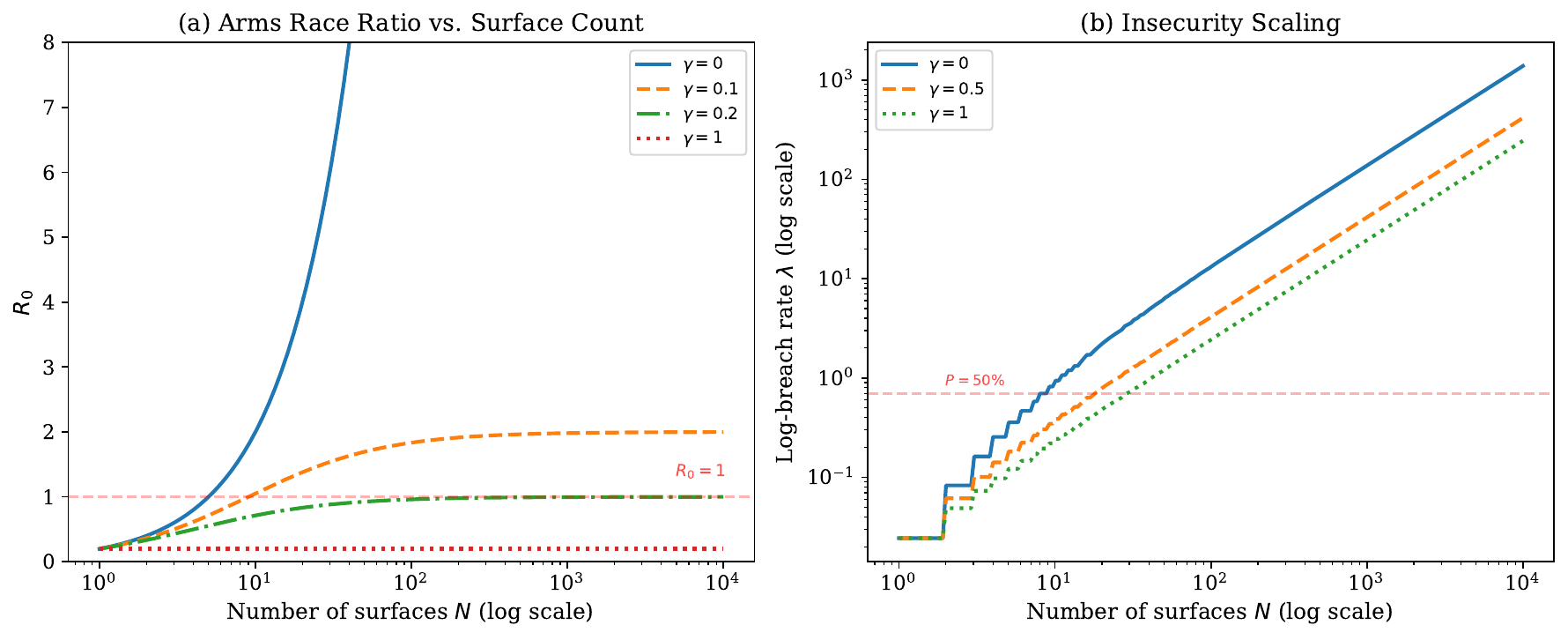}
\caption{(a)~Arms race ratio $R_0$ versus number of surfaces $N$ (log scale) for different signal cross-correlation levels $\gamma$ under full dilution ($\rho = 1$). At $\gamma = 1$, $R_0$ is flat (independent of $N$). At $\gamma = 0$, $R_0$ is linear in $N$, rapidly crossing the $R_0 = 1$ threshold. (b)~Log-breach rate $\lambda$ versus $N$ (log-log scale). Both regimes are asymptotically linear in $N$, but the coefficients differ dramatically: with signal cross-correlation, defense is effective; without it, per-surface defense effectiveness vanishes in the $N \to \infty$ limit under the dilution assumption.}
\label{fig:gamma}
\end{figure}

The $N$-independence result at $\rho = 1$, $\gamma = 1$ is exact in the contest form, not an approximation. It holds because cross-correlation restores every surface's effective signal to the full budget $s$, regardless of how many surfaces exist. Full cross-correlation removes the surface-count term from $R_0$: the condition for $R_0 < 1$ at the status quo is simply $\delta_0 s > \alpha$, which is achievable through signal capacity but not guaranteed by architecture alone. The $\gamma < 1$ case interpolates: partial cross-correlation slows the dilution but does not eliminate it, so $R_0$ grows with $N$ but more slowly than the $\gamma = 0$ baseline. Figure~\ref{fig:gamma} illustrates both effects.

\begin{figure}[H]
\centering
\includegraphics[width=0.75\textwidth]{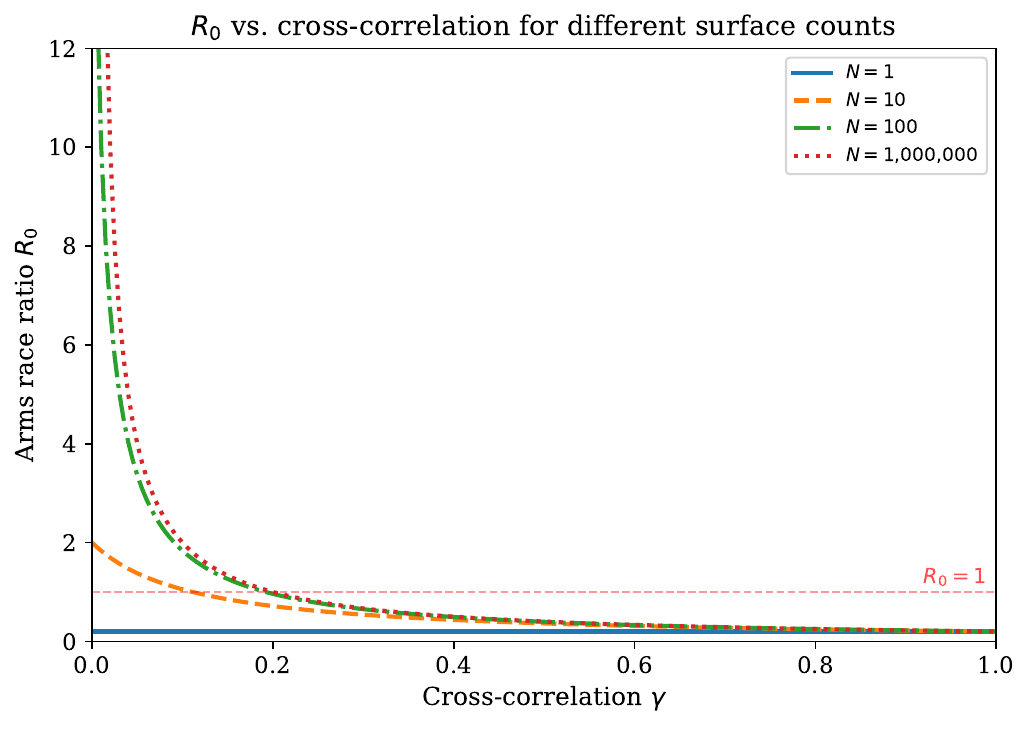}
\caption{Arms race ratio $R_0$ versus cross-correlation $\gamma$ for different surface counts $N$. At $N=1$, $\gamma$ does not appear in $R_0$ (flat line). For $N > 1$, increasing $\gamma$ sharply reduces $R_0$, with larger $N$ benefiting more. The dashed line marks $R_0 = 1$: above it, the marginal unit of attacker effort has a larger effect on breach probability; below, the defender's.}
\label{fig:R0_gamma}
\end{figure}

Figure~\ref{fig:R0_gamma} shows the same relationship from the defender's perspective. For a given surface count $N$, how much does $R_0$ improve as $\gamma$ increases? At $N = 1$, $\gamma$ is irrelevant because there is only one surface and nothing to cross-correlate. As $N$ grows, the gap between $\gamma = 0$ and $\gamma = 1$ widens dramatically, confirming that the marginal value of signal cross-correlation increases with the number of surfaces the defender must protect.

\begin{remark}[Marginal Value of Signal Cross-Correlation]\label{rem:dR0_dgamma}
The sensitivity of $R_0$ to signal cross-correlation at $\gamma = 0$ quantifies the marginal value of initial correlation investment. From equation~(\ref{eq:R0}):
\[
\left.\frac{\partial R_0}{\partial \gamma}\right|_{\gamma=0} = -\frac{\alpha \cdot N^\rho \cdot (N-1)}{\delta_0 \cdot s}
\]
The marginal benefit is proportional to $N^\rho(N-1)$: for $\rho = 1$ (full dilution), it is quadratic in $N$. At $N = 10$ surfaces, the marginal value of signal cross-correlation is nearly two orders of magnitude larger than at $N = 1$. This formalizes the intuition that signal cross-correlation becomes decisive at scale.
\end{remark}

\section{Extensions}
\label{sec:Extensions}

In this section, we step outside the formal model to explore its implications in richer strategic environments. We first consider the case in which $\gamma$ is endogenous, jointly determined by the complexity of the attacker's campaign and the defender's capacity to correlate signals across surfaces. We then analyze what happens when the attacker strategically selects its target from a heterogeneous population of defenders, drawing on the targeting framework of \citet{acemoglu2016network} but substituting the signal cross-correlation mechanism for network contagion.

\subsection{The Potency-Exposure Tradeoff}

Cross-correlation $\gamma$ has both a structural and a strategic component. Define $\gamma_a$ (structural cross-correlation created by the attack's multi-surface footprint within one defender's environment) and $\gamma_d$ (the defender's capacity to link signals across their own surfaces). Realized $\gamma = \min(\gamma_a, \gamma_d)$: the defender can only exploit cross-surface information that the attack's footprint actually generates.

\begin{center}
\begin{tabular}{p{3cm} p{5cm} p{5cm}}
\toprule
& \textbf{Low $\gamma_d$} & \textbf{High $\gamma_d$} \\
\midrule
\textbf{Simple attack} ($\gamma_a \approx 0$) & Status quo. & Wasted defender investment. \\[0.5em]
\textbf{Complex attack} ($\gamma_a$ high) & Attacker wins: high $h$, no information cost. & Ambiguous: high $h$ but high exposure. \\
\bottomrule
\end{tabular}
\end{center}

AI-augmented attacks tend toward greater complexity because AI excels at orchestrating multi-stage, multi-surface campaigns. Attacker AI adoption therefore endogenously raises $\gamma_a$, creating cross-surface dependencies that a correlating defender can exploit in detection. The attacker's investment in potency inadvertently generates the very information that funds the defender's correlation advantage.

The defender's investment in $\gamma_d$ thus functions as a commitment that shapes the attacker's strategy. High $\gamma_d$ makes complex attacks costly through increased exposure, pushing the attacker toward simpler, less potent strategies.

\begin{proposition}[Deterrence Threshold]\label{prop:deterrence}
Suppose the attacker can choose between a simple single-surface attack (amplification $h_s$, structural correlation $\gamma_a \approx 0$) and a complex multi-surface attack (amplification $h_c > h_s$, structural correlation $\gamma_a > 0$ from the multi-surface footprint). For a given defender signal breadth $s$ and defender correlation capacity $\gamma_d$, define the net benefit of complexity as:
\[
\Delta\pi(\gamma_d) = B \cdot [P_c(\gamma_d) - P_s] - c_a[a_c^* - a_s^*]
\]
where $P_c$ is the overall breach probability for the complex attack with realized correlation $\gamma = \min(\gamma_a, \gamma_d)$, and $P_s$ is the breach probability for the simple attack. Under the following sufficient conditions:
\begin{enumerate}[label=(\alph*)]
\item The complex attack spans $N_a \geq 2$ surfaces, creating structural correlation $\gamma_a > 0$.
\item The defender's correlation capacity $\gamma_d \in [0,1]$ is independent of the attack choice.
\item The simple attack targets a single surface with no cross-correlation benefit: $\gamma$ irrelevant.
\item The net benefit function $\Delta\pi$ is continuous and strictly decreasing on $[0,\gamma_a]$.
\item The endpoint signs satisfy $\Delta\pi(0) > 0 > \Delta\pi(\gamma_a)$.
\end{enumerate}
there exists a critical correlation capacity $\gamma_d^* \in (0, \gamma_a)$ such that:
\begin{enumerate}[label=(\roman*)]
\item For $\gamma_d > \gamma_d^*$: $\Delta\pi(\gamma_d) < 0$. The attacker's net benefit from complex multi-surface attacks is negative. The attacker optimally retreats to single-surface strategies, accepting lower potency to avoid information leakage. Correlation investment functions as a \emph{deterrent} against attack complexity.
\item For $\gamma_d < \gamma_d^*$: $\Delta\pi(\gamma_d) > 0$. Complex attacks are profitable. The attacker exploits the multi-surface structure with impunity.
\end{enumerate}
Moreover, for any parameter $x$ entering $\Delta\pi$, if $\partial \Delta\pi/\partial x$ exists and has a definite sign at the threshold, then
\[
\frac{d\gamma_d^*}{dx}
=
-\frac{\partial \Delta\pi/\partial x}{\partial \Delta\pi/\partial \gamma_d}.
\]
In particular, if $\partial \Delta\pi/\partial s < 0$, then $\gamma_d^*$ is decreasing in $s$; and if $\partial \Delta\pi/\partial h_c > 0$, then $\gamma_d^*$ is increasing in $h_c$.
\end{proposition}

\noindent Correlation capacity has a phase-transition character under this extension of the model. Below $\gamma_d^*$, investing in correlation achieves limited benefit because the attacker employs complex campaigns that partially overwhelm the defender's ability to correlate. Above $\gamma_d^*$, correlation deters complexity, pushing the attacker toward less potent strategies. This creates an economic argument for decisive rather than incremental correlation investment. The result extends the formal single-defender model by considering the attacker's choice between attack strategies, but unlike the convergence and scaling results, it is not derived from the base contest form alone.

\begin{remark}[Deterrence vs.\ Redirection for High-Stakes Adversaries] The deterrence threshold $\gamma_d^*$ has different practical content depending on the attacker's payoff $B$. When $B$ is very large relative to the cost of switching attack strategies---as in nation-state operations targeting high-value infrastructure---crossing above $\gamma_d^*$ does not induce the attacker to abandon the target. The participation constraint is not binding. Instead, $\gamma_d^*$ operates as a \emph{redirection} threshold: high $\gamma_d$ successfully eliminates complex, multi-surface campaigns, but the attacker substitutes toward simpler, lower-potency vectors that generate less cross-surface correlation and therefore evade the defender's correlation advantage. Signal correlation is still effective---it has driven attacker behavior toward less dangerous tactics---but targeted, high-payoff adversaries remain engaged. Proposition~\ref{prop:deterrence}'s ``retreat to single-surface strategies'' should be read as tactical simplification, not exit, when $B \gg c_a$.
\end{remark}

\subsection{Heterogeneous Defenders and the Targeting Externality}

The attacker faces a population of $K$ defenders, each characterized by $(d_k, s_k, \gamma_k, N_k, B_k)$. The attacker targets the defender with the highest expected value $B_k \cdot q_k$, which need not be the weakest in absolute terms: a moderate-vulnerability, high-value target such as a financial institution may be more attractive than a highly vulnerable but low-value one.

This targeting behavior creates a negative externality. Each defender's investment redirects attacker effort toward others, so the private return to defense exceeds the social return. The result is a zero-sum positioning game in which each defender over-invests to avoid being the marginal target. This is the beggar-thy-neighbor logic analyzed by \citet{acemoglu2016network}.

At the same time, there is underinvestment along a second margin. The formal model in Section~\ref{sec:signal} treats $\gamma$ as a within-organization parameter, capturing one defender's ability to link signals across its own $N$ surfaces. Individual defenders can raise this within-organization $\gamma$ through investment in SIEM, XDR, and cross-product detections. But a second, qualitatively different margin exists: cross-organization correlation, which involves detecting campaign-level patterns, novel TTPs, and coordinated threats visible only across many organizations' environments. No individual defender's investment can produce this; it requires aggregating signals from other defenders' environments.

Cross-organization correlation is a public good. Each defender who contributes signals to a shared detection infrastructure improves detection for all participants but captures only its own share of the benefit. Equilibrium provision of cross-organization $\gamma$ therefore falls below the social optimum by the standard public-good logic. Unlike the overinvestment result on private defense, which follows from the targeting externality, this underinvestment claim is an economic argument by analogy to the single-defender results rather than a derived equilibrium of a formal cross-organization game. The formal model establishes that $\gamma$ is the decisive variable; the public-good structure of cross-organization $\gamma$ is why the market undersupplies it.

Taken together, these two forces produce a dual inefficiency: simultaneous overinvestment in private defense (zero-sum positioning) and underinvestment in cross-organization signal correlation (positive-sum public good). A social planner would reallocate resources from the former to the latter.

\section{Conclusion}

The AI cybersecurity arms race is governed by six quantities: the attacker's AI potency ($\alpha$), the number of attack surfaces ($N$), the defender's AI effectiveness ($\delta_0$), signal breadth ($s$), cross-correlation ($\gamma$), and signal-processing returns to scale ($\rho$). These combine into the arms race ratio $R_0 = \alpha N^\rho / [\delta_0 s(1 + \gamma(N-1))]$, where $R_0 < 1$ favors the defender and $R_0 > 1$ favors the attacker.

The central result concerns the role of signal cross-correlation. With full cross-correlation ($\gamma = 1$) and full dilution ($\rho = 1$), the arms race ratio reduces to $R_0 = \alpha/(\delta_0 s)$, which is independent of $N$. The surface-count term drops out of $R_0$ entirely. When $\rho < 1$, the neutralization is partial, but $R_0$ decreases rather than increases in $N$, so the relative marginal effectiveness of defensive effort improves with scale. Within a single organization, this motivates unified detection infrastructure. Across organizations, it identifies the decisive margin: cross-organization signal correlation that no individual defender can produce.

Cross-organization signal correlation is a public good, and the market is likely to undersupply it while oversupplying private defense. This dual inefficiency, an economic argument extending the formal single-defender results rather than a derived multi-defender equilibrium, points to the structural importance of platform-scale signal aggregation. Which entities can resolve it will shape the evolution of the security industry.


\bibliographystyle{plainnat}
\bibliography{adversarial_discount}

\appendix

\section{Proofs and Derivations}

\subsection{Derivation of the Arms Race Ratio (Section~\ref{sec:arms_race})}

For per-surface breach probability $q = q_0 h(a)/[q_0 h(a) + (1-q_0)(1+\delta(a)ds_i^e)]$ in the contest form, the marginal effects are:
\begin{align*}
\frac{\partial q}{\partial a} &= \frac{q_0(1-q_0)[h'(a)(1+\delta(a)ds_i^e) + h(a)\lvert\delta'(a)\rvert ds_i^e]}{\Phi^2} \\
-\frac{\partial q}{\partial d} &= \frac{q_0(1-q_0)\,h(a)\,\delta(a)\,s_i^e}{\Phi^2}
\end{align*}
The $q_0(1-q_0)/\Phi^2$ factors cancel in the ratio $R = (\partial q/\partial a)/(-\partial q/\partial d)$, giving equation~(\ref{eq:Rgeneral}). At $(d,a)=(0,0)$: $h'(0) = \alpha$, $h(0) = 1$, $\delta(0) = \delta_0$, $\delta'(0)$ term vanishes (multiplied by $d=0$), giving $R_0 = \alpha/(\delta_0 s_i^e)$. With symmetric dilution and signal cross-correlation, $s_i^e = (s/N^\rho)(1+\gamma(N-1))$, yielding equation~(\ref{eq:R0}).

Since $\lambda = -N\log(1-q)$, $\partial\lambda/\partial a = N(\partial q/\partial a)/(1-q)$ and $\partial\lambda/\partial d = N(\partial q/\partial d)/(1-q)$. The $N/(1-q)$ factors cancel in the ratio, confirming $R$ is the same in $\lambda$ as in $q$.

\subsection{Proof of Proposition~\ref{prop:uniqueness} (Interior Equilibrium Uniqueness)}

At any interior defender optimum, the defender's FOC is $V \cdot q_0(1-q_0)h(a)\delta(a)s/\Phi^2 = c_d$, giving $\Phi^2 = Vq_0(1-q_0)h(a)\delta(a)s/c_d$. Since $\Phi = q_0h(a) + (1-q_0)(1+\delta(a)d s)$, this pins $\Phi$ as a function of $a$ alone along the interior branch. Solving for $d$:
\[
d^*(a) = \frac{\sqrt{Vq_0(1-q_0)h(a)\delta(a)s/c_d} - q_0 h(a) - (1-q_0)}{(1-q_0)\delta(a)s}
\]
Substituting into the attacker's payoff: $U_A = B\sqrt{q_0 c_d h(a)/(V(1-q_0)\delta(a)s)} - c_a a$. This is strictly concave in $a$ iff $f(a) = \sqrt{h(a)/\delta(a)}$ is strictly concave.

\textit{Concavity condition.} Define $\varphi = \log h - \log \delta$, so $f = e^{\varphi/2}$. Then $f'' = (\varphi''/2 + \varphi'^2/4)e^{\varphi/2}$, so $f'' < 0$ iff $\varphi'' + \varphi'^2/2 < 0$. Computing:
\[
\varphi' = \frac{h'}{h} + \frac{\lvert\delta'\rvert}{\delta}, \qquad \varphi'' = \frac{h''}{h} - \left(\frac{h'}{h}\right)^2 - \frac{\delta''}{\delta} + \left(\frac{\lvert\delta'\rvert}{\delta}\right)^2
\]
Substituting and collecting:
\[
f'' < 0 \iff \frac{h''}{h} - \frac{1}{2}\left(\frac{h'}{h} - \frac{\lvert\delta'\rvert}{\delta}\right)^2 < \frac{\delta''}{\delta} - 2\left(\frac{\delta'}{\delta}\right)^2
\]
The LHS is strictly negative when $h'' < 0$. The RHS equals $\delta \cdot \delta'' - 2(\delta')^2$ divided by $\delta^2$, which is non-negative iff $\delta \cdot \delta'' \geq 2(\delta')^2$.

\textit{Hyperbolic erosion.} For $\delta(a) = \delta_0/(1+\beta a)$: $\lvert\delta'\rvert/\delta = \beta/(1+\beta a)$ and $\delta''/\delta = 2\beta^2/(1+\beta a)^2 = 2(\lvert\delta'\rvert/\delta)^2$. The RHS is exactly zero, and the condition reduces to $h''/h < (1/2)(h'/h - \lvert\delta'\rvert/\delta)^2$, which holds whenever $h'' < 0$ since the LHS is negative and the RHS is non-negative.

Thus the reduced attacker problem is strictly concave along the interior branch and admits at most one maximizer. This establishes that there is at most one interior equilibrium. Corner equilibria, if any, are not covered by this argument because the defender's constrained best response is $d^{BR}(a) = \max\{0, d^*(a)\}$ rather than the interior formula alone.

\subsection{Proof of Proposition~\ref{prop:stability} (Global Convergence)}

\textit{Bounded best responses.} The constrained best responses are bounded directly from the payoff ranges. Since $q \in [0,1]$, the defender's payoff satisfies $U_D(d,a) = -Vq(d,a) - c_d d \le -c_d d$. At $d=0$, the defender obtains $U_D(0,a) = -Vq(0,a) \ge -V$. Hence any $d > V/c_d$ yields $U_D(d,a) < -V \le U_D(0,a)$ and cannot be optimal, so $d^{BR}(a) \le V/c_d$ for all $a$. Similarly, the attacker's payoff satisfies $U_A(d,a) = Bq(d,a) - c_a a - F\mathbf{1}(a>0) \le B - c_a a$. At $a=0$, the attacker obtains $U_A(d,0) = Bq(d,0) \ge 0$. Hence any $a > B/c_a$ yields $U_A(d,a) < 0 \le U_A(d,0)$ and cannot be optimal, so $a^{BR}(d) \le B/c_a$ for all $d$. Both best responses are therefore contained in a compact set $[0, V/c_d] \times [0, B/c_a]$, which is forward-invariant under the dynamics.

\textit{No closed orbits (Bendixson).} The vector field $F = (d^{BR}(a) - d,\; a^{BR}(d) - a)$ has divergence $\text{div}(F) = -1 + (-1) = -2$ on each smooth region, because $d^{BR}(a)$ does not depend on $d$ and $a^{BR}(d)$ does not depend on $a$. By Green's theorem, any closed invariant curve $\Gamma$ contained in a smooth region and enclosing region $\Omega$ would require $\oint_\Gamma F \cdot n\, ds = \iint_\Omega \text{div}(F)\, dA = -2 \cdot \text{Area}(\Omega) < 0$, contradicting the zero net flux through an invariant curve. This rules out periodic orbits and homoclinic loops within each smooth region.

\textit{Corner solutions.} At corner solutions where $d^{BR} = 0$ or $a^{BR} = 0$, the vector field $F$ is piecewise-$C^1$. The boundary $d = 0$ is absorbing from below (cannot go negative) and the dynamics on this boundary reduce to one-dimensional flow in $a$. Similarly for $a = 0$. The divergence argument applies separately to each smooth region, and the Bendixson criterion extends by considering the net flux across the piecewise-smooth boundaries. Since both constrained best responses satisfy $d^{BR}(a) \geq 0$ and $a^{BR}(d) \geq 0$, trajectories cannot escape the positive orthant. For a rigorous treatment of Poincaré-Bendixson on piecewise-smooth planar systems, see \citet{filippov1988differential}.

\textit{Saddle exclusion.} The Jacobian of $F$ at the equilibrium is $J = \bigl[\begin{smallmatrix} -1 & \partial d^{BR}/\partial a \\ \partial a^{BR}/\partial d & -1 \end{smallmatrix}\bigr]$, with $\text{tr}(J) = -2$ and $\det(J) = 1 - (\partial d^{BR}/\partial a)(\partial a^{BR}/\partial d)$. A saddle requires $\det(J) < 0$, implying a 1D unstable manifold. Each branch of the unstable manifold is bounded, cannot approach a periodic orbit (Bendixson), and cannot approach another fixed point (unique). By the Poincar\'{e}-Bendixson theorem, it must return to the saddle, forming a homoclinic loop---which the area argument above forbids. Therefore $\det(J) > 0$: the equilibrium is a stable node or spiral.

\textit{Global convergence (Poincar\'{e}-Bendixson).} In two dimensions, the Poincar\'{e}-Bendixson theorem states that a bounded trajectory that does not approach a closed orbit must converge to a fixed point. With bounded trajectories, no closed orbits or homoclinic loops, and a unique stable equilibrium, every trajectory converges to that equilibrium.

\textit{Discrete-time extension.} For the discrete system with adjustment rate $\eta > 0$, the linearized Jacobian at the equilibrium is $J_\eta = (1-\eta)I + \eta J_0$, where $J_0$ has eigenvalues $\pm\sqrt{\partial d^{BR}/\partial a \cdot \partial a^{BR}/\partial d}$. Local stability holds when $\eta < 2/(\rho_0 + 1)$, where $\rho_0 = \sqrt{\lvert\partial d^{BR}/\partial a \cdot \partial a^{BR}/\partial d\rvert}$ is the undamped spectral radius. Numerical simulation confirms global convergence for $\eta = 0.15$ across $\beta \in [0.1, 10]$, $B/V \in [0.125, 3]$.

\subsection{Derivation of Proposition~\ref{prop:scaling} (Surface Scaling)}

\textit{Part (i).} With $\gamma = 1$ and dilution parameter $\rho$: $s_i^e = (s/N^\rho)(1 + 1\cdot(N-1)) = sN^{1-\rho}$. Substituting into $R_0 = \alpha N^\rho/(\delta_0 s_i^e)$ gives $R_0 = \alpha N^\rho/(\delta_0 s N^{1-\rho}) = \alpha N^{\rho-1}/(\delta_0 s)$. At $\rho = 1$: $R_0 = \alpha/(\delta_0 s)$, independent of $N$. At $\rho < 1$: $R_0 \propto N^{\rho-1}$ decreases in $N$. At $\rho = 0$: $R_0 = \alpha/(N \delta_0 s)$, strongly decreasing in $N$.

\textit{Part (ii).} With $\gamma = 0$: $s_i^e = s/N^\rho$. Per-surface breach probability:
\[
q_i = \frac{q_0 h(a)}{q_0 h(a) + (1-q_0)(1 + \delta(a)ds/N^\rho)}
\]
As $N \to \infty$ with $\rho > 0$, the defense term $\delta(a)ds/N^\rho \to 0$, so $q_i \to q_\infty \equiv q_0 h(a)/[q_0 h(a) + (1-q_0)]$, which contains no defense terms. The log-breach rate $\lambda = -N\log(1-q_i) \to -N\log(1-q_\infty)$, which is linear in $N$ at a coefficient reflecting zero per-surface defense. At $\rho = 0$ (no dilution), $s_i^e = s$ for all $N$, so defense does not vanish and $q_i$ remains bounded away from $q_\infty$.

\textit{Part (iii).} A critical $N^*$, when it exists, satisfies $R_0(N^*) = 1$, i.e.
\[
\alpha N^{*\rho} = \delta_0 s [1 + \gamma(N^*-1)].
\]
Equivalently, define
\[
F(N,\gamma) \equiv \alpha N^\rho - \delta_0 s [1 + \gamma(N-1)].
\]
Then $N^*$ solves $F(N^*,\gamma)=0$. If $\partial F/\partial N \neq 0$ at the root, the implicit function theorem yields a locally unique threshold and
\[
\frac{dN^*}{d\gamma}
=
-\frac{\partial F/\partial \gamma}{\partial F/\partial N}
=
\frac{\delta_0 s\,(N^*-1)}{\alpha \rho (N^*)^{\rho-1} - \delta_0 s\,\gamma}.
\]
Hence $N^*$ is increasing in $\gamma$ whenever $\alpha \rho (N^*)^{\rho-1} > \delta_0 s\,\gamma$. At $\gamma = 1$, the equation simplifies to $N^{*\rho} = (\delta_0 s / \alpha) N^*$, giving $N^{*\rho-1} = \delta_0 s / \alpha$. For $\rho = 1$, this yields $N^*$-independence: the parity condition becomes $\alpha = \delta_0 s$, independent of surface count.

\subsection{Proof of Proposition~\ref{prop:deterrence} (Deterrence Threshold)}

The proposition extends the base model by allowing the attacker to choose between attack strategies. The proof establishes existence, uniqueness, and conditional comparative statics for the threshold.

The attacker chooses between a complex multi-surface attack ($N_a$ surfaces, high $h_c$, structural correlation $\gamma_a > 0$ from the multi-surface footprint) and a simple single-surface attack (1 surface, lower $h_s$, no structural correlation). For the complex attack, realized $\gamma = \min(\gamma_a, \gamma_d)$; effective per-surface signal $s_i^e = (s/N_a^\rho)[1 + \gamma(N_a - 1)]$; overall breach probability $P_c = 1 - \prod_{i=1}^{N_a}(1-q_i)$ where each $q_i$ depends on $s_i^e$. For the simple attack: single surface with $\gamma$ irrelevant (only one surface), $q_s = q_0 h_s/(q_0 h_s + (1-q_0)(1+\delta_s d s))$.

Define net benefit: $\Delta\pi(\gamma_d) = B[P_c(\gamma_d) - P_s] - c_a[a_c^* - a_s^*]$.

\textit{Monotonicity.} For $\gamma_d \in [0, \gamma_a]$, realized $\gamma = \gamma_d$ increases with the defender's capacity. Higher $\gamma_d$ raises $s_i^e$ for the complex attack, reducing each $q_i$ and therefore $P_c$. Since the simple attack is unaffected, $\partial P_c/\partial \gamma_d < 0$ implies $\partial \Delta\pi/\partial \gamma_d < 0$: $\Delta\pi$ is strictly decreasing in $\gamma_d$ on $[0, \gamma_a]$. For $\gamma_d > \gamma_a$, realized $\gamma = \gamma_a$ is constant, so $\Delta\pi$ is flat.

\textit{Boundary conditions.} At $\gamma_d = 0$, realized $\gamma = 0$, so the complex attack faces maximum dilution and each surface sees $s_i^e = s/N_a^\rho$. At $\gamma_d = \gamma_a$, realized $\gamma = \gamma_a$, so
\[
s_i^e = \frac{s}{N_a^\rho}[1 + \gamma_a(N_a-1)].
\]
At $\rho = 1$, this simplifies to
\[
s_i^e = \frac{s}{N_a}[1 + \gamma_a(N_a-1)]
= s\left(\gamma_a + \frac{1-\gamma_a}{N_a}\right),
\]
which equals $s$ when $\gamma_a = 1$. By assumption, $\Delta\pi(0) > 0 > \Delta\pi(\gamma_a)$.

\textit{Existence and uniqueness of the threshold.} By continuity of $\Delta\pi$ on $[0, \gamma_a]$, strict monotonicity, and opposite signs at the boundaries, the intermediate value theorem guarantees existence of $\gamma_d^* \in (0, \gamma_a)$ where $\Delta\pi(\gamma_d^*) = 0$. Strict monotonicity implies that this root is unique.

\textit{Comparative statics.} Let $x$ denote any parameter entering $\Delta\pi$. Implicit differentiation of $\Delta\pi(\gamma_d^*;x) = 0$ gives
\[
\frac{d\gamma_d^*}{dx}
=
-\frac{\partial \Delta\pi/\partial x}{\partial \Delta\pi/\partial \gamma_d}.
\]
Since $\partial \Delta\pi/\partial \gamma_d < 0$, the sign of $d\gamma_d^*/dx$ matches the sign of $\partial \Delta\pi/\partial x$. Therefore, if $\partial \Delta\pi/\partial s < 0$, then $d\gamma_d^*/ds < 0$; and if $\partial \Delta\pi/\partial h_c > 0$, then $d\gamma_d^*/dh_c > 0$.

\end{document}